\documentclass{XrU2005}
\usepackage{graphicx}

\title{The High-Velocity Outflow of PG 1211+143: An Unbiased View Based on Several Observations}
\author[1,2]{S. Kaspi}
\author[1]{E. Behar}
\affil[1]{Department of Physics, Technion, Haifa 32000, Israel}
\affil[2]{School of Physics and Astronomy and the Wise
Observatory, Tel-Aviv University, Tel-Aviv 69978, Israel}

\newcommand  \kms      {\ifmmode {\rm km\,s}^{-1} \else km\,s$^{-1}$\fi}
\newcommand  \cmii     {\hbox{cm$^{-2}$}}
\newcommand  \ergs     {\ifmmode {\rm erg\,s}^{-1} \else erg s$^{-1}$\fi}
\newcommand  \ergcms   {\ifmmode {\rm erg\,cm}^{-2}\,{\rm s}^{-1}
                        \else erg\,cm$^{-2}$\,s$^{-1}$\fi}
\newcommand  \ergcmsA  {\ifmmode{\rm erg\,cm}^{-2}\,{\rm s}^{-1}\,{\rm\AA}^{-1}
                        \else erg\,cm$^{-2}$\,s$^{-1}$\,\AA$^{-1}$\fi}
\newcommand  \ergcmsHz {\ifmmode{\rm erg\,cm}^{-2}\,{\rm s}^{-1}\,{\rm Hz}^{-1}
                        \else erg\,cm$^{-2}$\,s$^{-1}$\,Hz$^{-1}$\fi}
\newcommand  \phcms    {\ifmmode {\rm ph\,cm}^{-2}\,{\rm s}^{-1}
                        \else ph\,cm$^{-2}$\,s$^{-1}$\fi}
\newcommand  \phcmsA   {\ifmmode {\rm ph\,cm}^{-2}\,{\rm s}^{-1}\,{\rm\AA}^{-1}
                        \else ph\,cm$^{-2}$\,s$^{-1}$\,\AA$^{-1}$\fi}

\newcommand  \pgt      {PG\,1211+143}
\newcommand  \xmm      {{\it XMM-Newton}}

\begin{document}


\maketitle

\begin{abstract}
We present and discuss high-resolution grating spectra of the
quasar \pgt\ obtained over three years.  Based on an early
observation from 2001, we find an outflow component of about 3000
\kms\ in contrast with the much higher velocity of about 24000
\kms\ reported earlier for this source, and based on the same data
set. Subsequent grating spectra obtained for \pgt\ are consistent
with the first observation in the broad-band sense, but not all
narrow features used to identify the outflow are reproduced. We
demonstrate that the poor S/N and time variability seen during all
existing observations of \pgt\ make any claims about the outflow
precariously inconclusive.
\end{abstract}

\section{Introduction}

Typical mass outflow velocities of a few hundreds to a few thousands
\kms\ have been measured by now in numerous Active Galactic Nuclei
(AGNs; Crenshaw et al. 2003 and references therein). Recent studies
of the X-ray spectra of certain quasars have led to claims of much
higher outflow velocities reaching a significant fraction of the
speed of light, e.g., APM\,08279+5255 --- Chartas et al. (2002)
claim speeds of $\sim0.2c$ and $\sim0.4c$\,\footnote{Though Hasinger
et al.  (2002) using a different instrument prefer a more conservative
interpretation by which the X-ray wind is much slower and consistent
with the well known, UV broad absorption line wind of that source,
outflowing at velocities of up to 12000 \kms .}, PG\,1115+080 ---
Chartas et al. (2003) find two X-ray absorption systems with outflow
velocities of $\sim0.10c$ and $\sim0.34c$.  These measurements,
however, were carried out using spectra obtained with CCD cameras
and hence at moderate spectral resolving powers of $R\sim 50$. Using
the \xmm\ reflection gratings ($R$ up to 500), high resolution
X-ray spectra for several quasars have been obtained. For \pgt\
Pounds et~al. (2003a, 2005) find a rich, well resolved spectrum featuring
absorption lines of several ions, which they interpret as due to an
outflow of $\sim$~24000~\kms. A similar interpretation was applied
to a similar observation of PG\,0844+349, where Pounds et al. (2003b)
report even higher velocities reaching $\sim$~60000~\kms. In NGC\,4051,
Pounds et al. (2004) find a single absorption line at $\sim$~7.1~keV,
which they suggest may be {Fe}\,{\sc xxvi} Ly$\alpha$ at an
outflow velocity of $\sim$~ 6500~\kms, or the He$\alpha$ resonance
absorption line of {Fe}\,{xxv} in which case the outflow velocity
is $\sim$~16500~\kms. Yet another ultra-high-velocity  (UHV, i.e.,
sub-$c$) wind of 50000~\kms\ was reported by Reeves et al. (2003)
for PDS~456. In all of these sources, the inferred hydrogen column
densities through the wind is of the order of 10$^{23}$~\cmii, which
is about an order of magnitude higher than the typical values measured
for the nearby Seyfert sources.

If indeed UHV outflows are common to bright quasars, this could have
far reaching implications on our understanding of AGN winds and AGNs
in general. For instance, if these winds carry a significant amount
of mass as the high column densities may suggest, they would alter
our estimates of the metal enrichment of the intergalactic medium by
quasars. It remains to be shown theoretically what mechanism (e.g.,
radiation pressure) can drive these intense winds. Since the amount
of mass in the wind is not well constrained, it is still unclear what
effect it may have on the energy budget of the AGN.  King \& Pounds
(2003) note that UHV winds have been found mostly for AGNs accreting
near their Eddington limit. They provide a theory by which the UHV
outflows are optically thick producing an effective photosphere,
which is also responsible for the UV blackbody and soft X-ray (excess)
continuum emission observed for these sources.

\section{\pgt\ \ --- \ first observation \ --- \ second view}

\begin{figure}
\centering
\centerline{\includegraphics[width=7.5cm]{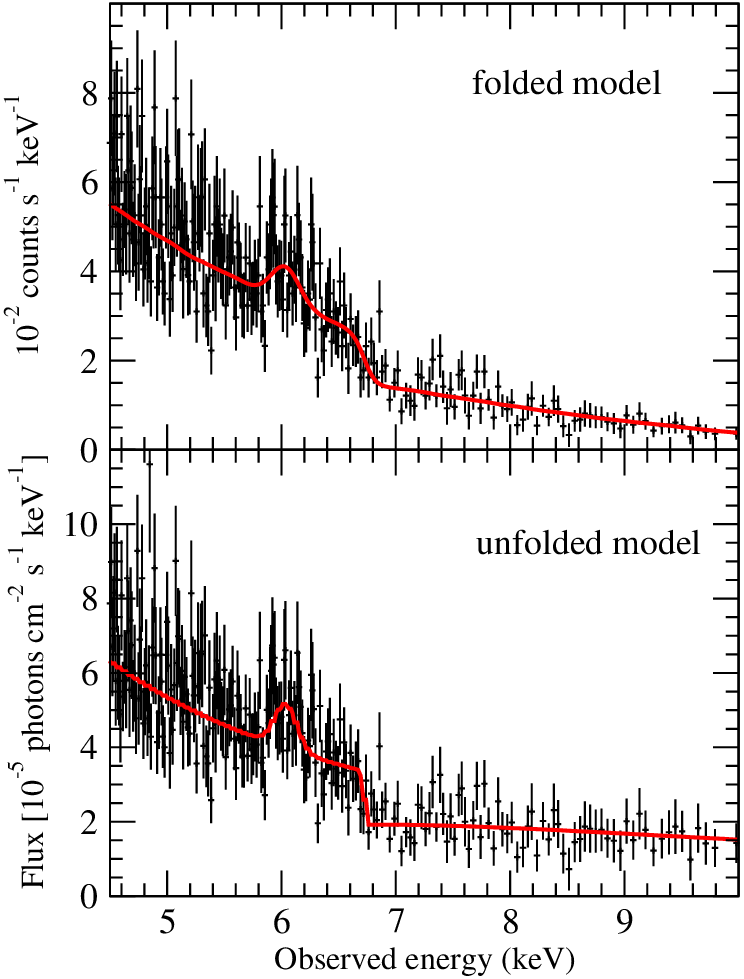}}
\caption{The EPIC-pn data and a simple fitted model, which is discussed
in \S~\ref{epicpn}. The upper panel shows the model folded through the
instrument response and compared with the data.  Bottom panel shows
the unfolded model.}
\label{pnmodel}
\end{figure}

\pgt\ was observed with {\it XMM-Newton} during 2001 June 15 for
about 55 ks. We retrieved the data for this observation from the
{\it XMM-Newton} archive and reduced them using the Science Analysis
System (SAS v5.3.0) in the standard processing chains as described in
the data analysis threads and the ABC Guide to {\it XMM-Newton} Data
Analysis. Overall, our data reduction results agree well with that of
Pounds et al. (2003a, 2005) except for a few minor features which appear
to be slightly different between the two reductions.  We attribute
these discrepancies to the different binning methods used and to the
averaging of RGS1 and RGS2 in this work (see below). The results
described in this section are presented in details in Kaspi \& Behar
(2006).

\subsection{EPIC-pn}
\label{epicpn}

For the EPIC-pn data we first fitted the (line-free) rest-frame 2--5
keV energy range with a simple power law. The best fitted power law
has a photon index of $\Gamma = 1.55\pm 0.05$ and a normalization
of $(6.6\pm 0.4)\times10^{-4}$ ph\,cm$^{-2}$\,s$^{-1}$\,keV$^{-1}$
and gives $\chi^{2}_{\nu}=0.74$ for 487 degrees-of-freedom
(d.o.f.). Extrapolating this power law up to a rest-frame energy of
11~keV, we find a flux excess above the power law at around 6.4~keV,
which is indicative of an iron K$\alpha$ line, and a flux deficit
below the power law at energies above 7~keV.  We add to the model
a Gaussian emission line to account for the Fe\,K$\alpha$ line and
a photoelectric absorption edge to account for the deficit. Fitting
for all parameters simultaneously, we find the best-fit Gaussian line
center is at $6.04\pm 0.04$~keV (or $6.53\pm 0.05$~keV in the rest
frame) and a line width ($\sigma$) of $0.096\pm 0.067$~keV. The total
flux in the line is $(2.9\pm 1.4)\times 10^{-6}$ \phcms.  For the edge,
we find a threshold energy of $6.72\pm 0.10$~keV, which is translated
to a rest frame energy of $7.27\pm 0.11$~keV. The optical depth at the
edge is $\tau=0.56\pm 0.10$. The power law model with the Gaussian
line and the absorption edge gives a $\chi^{2}_{\nu}=0.983$ for 613
d.o.f. This model is plotted in Figure~\ref{pnmodel} where we show the
model, both folded through the instrument and fluxed (i.e., unfolded).
We stress that this edge does not necessarily contradict the presence
of the line detected by Pounds et al. (2003a, 2005), since K$\alpha$ edges
have lines right next to them.

We also observe the lines at 2.68 keV and 1.47 keV claimed by Pounds
et al. (2003a, 2005) to be from S and Mg, only we identify them as different
lines at much lower velocities. The 2.68 keV line is identified here
as {S}\,{\sc xv}\,He$\beta$ and the 1.47 keV line is identified as
{Mg}\,{\sc xi}\,He$\beta$.

\subsection{RGS}

\begin{figure*}
\centering
\includegraphics[width=16cm]{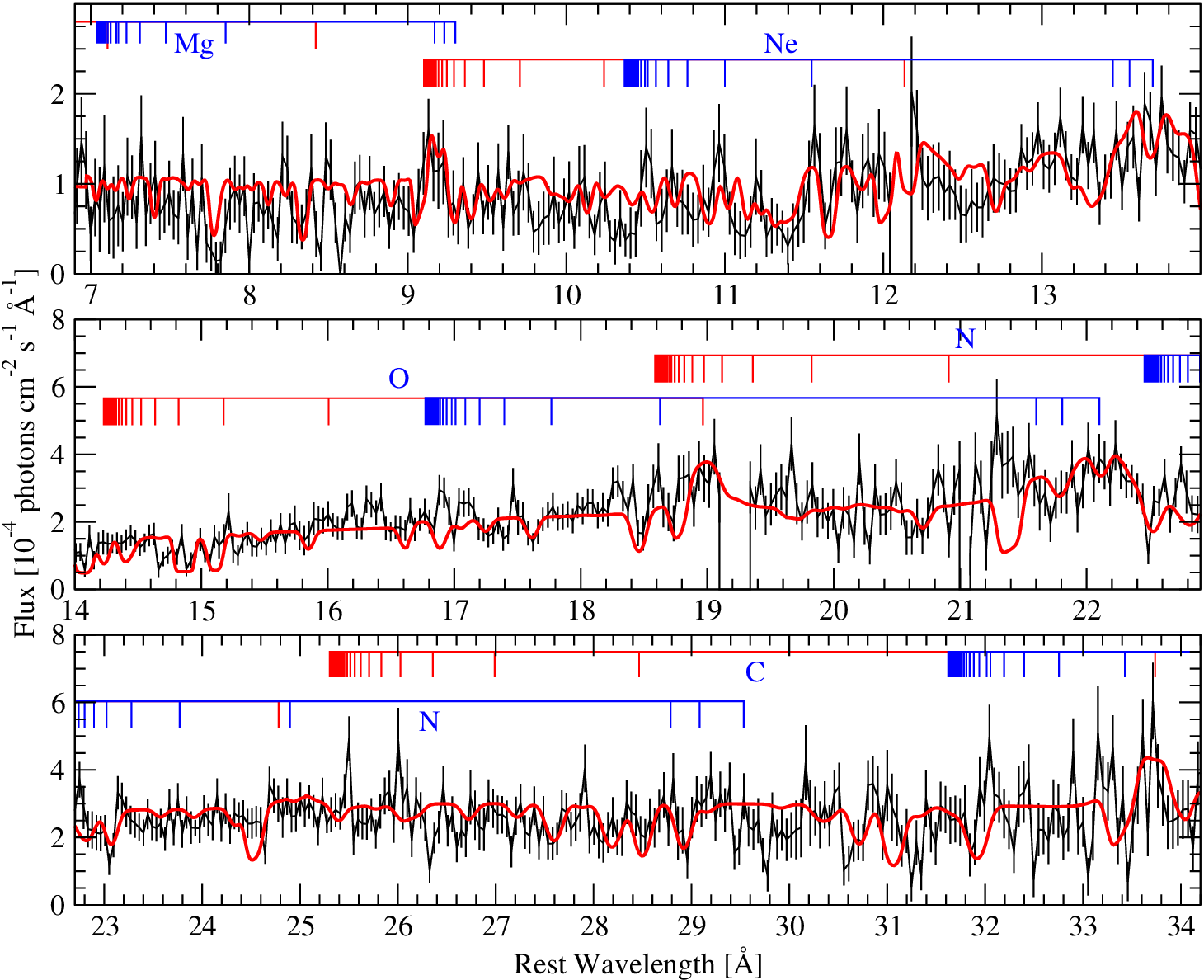}
\caption{Combined RGS1 and RGS2 spectrum of \pgt\ binned to $\sim 0.04$
\AA. The spectrum has been corrected for Galactic absorption and for
the redshift of the source. The rest-frame positions of lines from
H-like and He-like ions of N, O, Ne, and Mg are marked
above the spectrum. The lines from the lower ionization states of
O and Mg, and the L-shell lines of Fe are not marked. Gaps in the
spectrum are due to chip gaps and have zero flux. The model
is marked as the solid red curve.}
\label{pgtrgs}
\end{figure*}

The RGS1 and RGS2 were operated in the standard spectroscopy mode
resulting in an exposure time of $\sim 52$ ks.  The spectra were
extracted into uniform bins of $\sim$\,0.04 \AA\ (which is about
the RGS resolution and is 4 times the default bin width) in order to
increase the signal-to-noise ratio (S/N). For the purpose of modeling
narrow absorption lines, this rebinning method is better than the
method used by Pounds et al. (2003a) of rebinning the spectrum to a
minimum of 20 counts per bin, which distorts the spectrum especially
around low-count-rate absorption troughs. To flux calibrate the
RGS spectra we divided the count spectrum of each instrument by
its exposure time and its effective area at each wavelength. Each
flux-calibrated spectrum was corrected for Galactic absorption and the
two spectra were combined into an error-weighted mean. At wavelengths
where the RGS2 bins did not match exactly the wavelength of the
RGS1 bins, we interpolated the RGS2 data to enable the averaging.
The sky-subtracted combined RGS spectrum has in total $\sim 8900$
counts and its S/N ranges from $\sim 2$ around 8\,\AA\ to $\sim 5$
around 18\,\AA\ with an average of 3.  Statistics in the second
order of refraction are insufficient, hence we did not include it in
our analysis.

The combined RGS spectrum (RGS1 and RGS2) of \pgt\ is presented in
Figure~\ref{pgtrgs}. Numerous absorption lines and several emission
lines are detected.  We identify K-shell lines of C, N, O and Mg and
L-shell lines of O, Mg, Si, Ar, and Fe.  The absorption line widths
are consistent with the RGS resolution, and with the present S/N we
are not able to resolve the intrinsic velocity widths.  In emission,
we identify significantly broadened lines of {N}\,{\sc vii} Ly$\alpha$,
{O}\,{\sc viii} Ly$\alpha$, the forbidden line of {O}\,{\sc vii} and
its He$\alpha$ resonance line, the forbidden line of {Ne}\,{\sc ix},
and the {Mg}\,{\sc xi} He$\alpha$ resonance line, all in the rest
frame of the source with no velocity shift.

In order to quantitatively explore the emission and absorption lines,
we have constructed a model for the entire RGS spectrum.  The present
method is an ion-by-ion fit to the data similar to the approach used
in Sako et al. (2001) and in Behar et~al. (2003). We first use the
continuum measured from the EPIC-pn data, but renormalized to the
RGS flux level. This continuum is then absorbed using the full set
of lines for each individual ion. Our absorption model includes
the first 10 resonance lines of H- and He- like ions of C, N, O,
Ne, and Mg as well as edges for these ions. The model also includes
our own calculation for the L-shell absorption lines of Fe (Behar et
al. 2001) as well as of Si, S, and Ar corrected according to laboratory
measurements (Lepson et al. 2003, 2005). Finally, we include inner-shell
K$\alpha$ absorption lines of O and Mg (Behar \& Netzer 2002), which
we detect in the spectrum. The absorbed spectrum is complemented by
the emission lines mentioned above, which are observed in the RGS
spectrum.

By experimenting with the absorption line parameters, we find that
the observed lines are all blueshifted by about 3000~\kms\ with an
uncertainty of 500~\kms . In the model we used a turbulence velocity
of 1000~\kms\ to broaden the absorption lines.  This width includes
the instrumental broadening, which as noted above, we could not
separate from the intrinsic broadening.  Since the lines appear to
be saturated, but no line goes to zero intensity in the trough, we
obtain the best fit by assuming a covering factor of 0.7 for the X-ray
continuum source. The best-fit column densities that we find for the
different ions are consistent with a hydrogen column density of about
$10^{21}$--$10^{22}$ \cmii .  The emission lines are modeled using
Gaussians with uniform widths of $\sigma = 2500$ \kms\ (resolved, but
again, including the instrumental broadening), with no velocity shift,
and assumed to be unabsorbed. These lines at FWHM~$\simeq$ $6000\pm
1200$~\kms\ are even broader than those observed from the broad line
region in the visible band ($\sim 2000$ \kms ; Kaspi et al. 2000). The
entire best-fit spectrum is shown in Figure~\ref{pgtrgs} (red curve).
The spectrum beyond 25~\AA\ is particularly challenging as it comprises
many unresolved lines from L-shell ions of Si, S, and Ar while the
RGS effective area drops rapidly. Several predicted lines may be
observed here (e.g., {Ar}\,{\sc xiii} - 28.92~\AA, {Si}\,{\sc xii} -
30.71~\AA, {Ar}\,{\sc xii} - 31.06~\AA, {S}\,{\sc xiii} - 31.93~\AA;
these wavelengths include the 3000 \kms\ shift). We are still unable to
explain several features seen in the data, e.g., around 8.5 \AA, 10.4
\AA, or 29.8 \AA , but the model gives a good fit to the data overall.

\subsection{Conclusions - First Observation}

We have provided a self consistent model to the ionized outflow
of \pgt\ revealing an outflow velocity of approximately 3000~\kms.
Our model reproduces many absorption lines in the RGS band, although
the S/N of the present data set is rather poor and some of the noise
might be confused with absorption lines.

The present approach is distinct from the commonly used global
fitting methods and also from the line-by-line approach used by
Pounds et al. (2003a). It allows for a physically consistent
fit to the spectrum and is particularly appropriate for a
broad-ionization-distribution absorber as observed here for \pgt .

The present model also features several broad (FWHM = 6000~\kms)
emission lines, which are observed directly in the data.

A broad and relatively flat ionization distribution is found
throughout the X-ray outflow consistent with a hydrogen column density
of roughly $10^{21}$--$10^{22}$ \cmii. This is reminiscent of the outflow
parameters measured in other well studied Seyfert galaxies.

We also detect Fe-K absorption, which was identified by Pounds
et~al. (2003a, 2005) as a strongly blueshifted {Fe}\,{\sc xxvi} absorption
line. We find that most of the Fe-K opacity can alternatively be
attributed to several consecutive, low charge states of Fe, although
it can not be assessed whether the absorber is co-moving with the
outflow or not. Future missions with microcalorimeter spectrometers
on board might be able to address this interesting question.

\section{A second Observation of \pgt}

\begin{figure*}
\centering
\includegraphics[width=16.2cm]{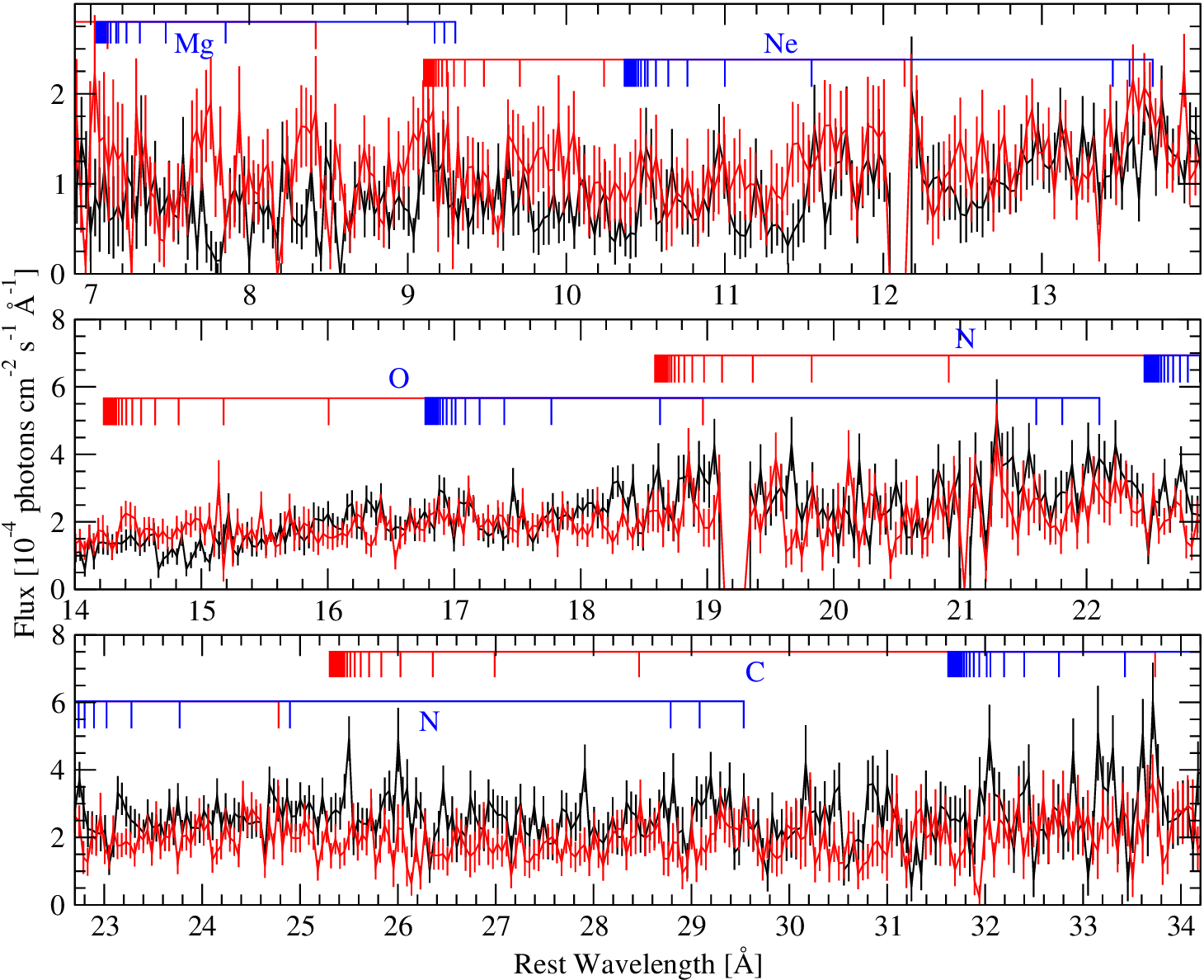}
\caption{Two RGS spectra of \pgt\ obtained three years apart. The
first observation (in black) was carried out on 2001 June 15 and
is also presented in Figure 1. The second observation (in red)
took place on 2004 June 21.} \label{2rgs}
\end{figure*}

A second \xmm\ observation of \pgt\ for $\sim 50$\,ks was carried
out on 2004 June 21, three years after the first observation of 2001
June 15 which is described above. We have retrieved the data of this
second observation from the \xmm\ archive and reduced it in exactly the
same way described above for the first observation.  The data of the
second RGS observation are plotted (red line) in Figure~\ref{2rgs}
over the first observation (black line). The broad-band spectra
of the two observations are generally consistent. However, not all
narrow features are consistently reproduced. The total flux in the
RGS band is the same in the two observations, though the continuum
slope in the second observation the spectrum is somewhat harder.

When inspecting the detailed narrow features in the spectrum some
have changed while others remain the same. For example, the second
observations shows features which appear to be emission lines around
8 \AA\, where the first observation had absorption lines.  Also,
around 15 \AA\ the absorption lines seem to have disappeared in the
second observation. Conversely, some features are the same in the two
spectra, for example, the emission O\,{\sc vii} triplet around 22 \AA\
and the Ne\,{\sc ix} triplet around 13.5 \AA.  From Figure~3, it can
be seen that due to the poor S/N in both spectra, it is extremely hard
to determine whether the differences between the two spectra are real,
or a mere result of the data's poor S/N.

\section{Simultaneous observations of \xmm\ and {\it Chandra}}

\begin{figure*}
\centering
\includegraphics[width=17cm]{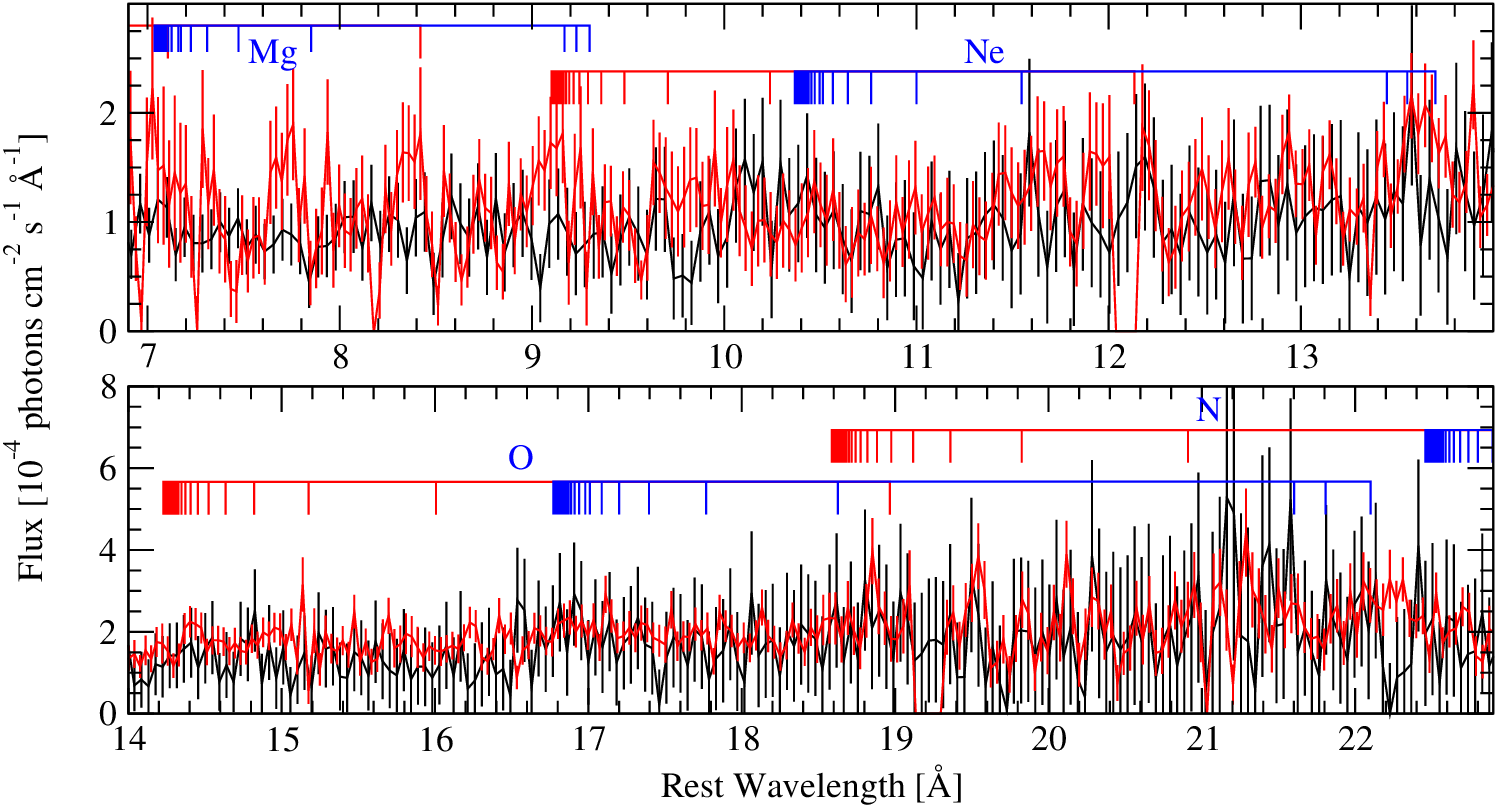}
\caption{ \pgt\ spectrum taken by the RGS on board \xmm\ (red line)
on 2004 June 21 shown together with a spectrum taken simultaneously
by the LETG on board {\it Chandra} (black line). Although the data are
taken at the same time, discrepancies between the spectra are evident,
probably as a result of the poor S/N of the data.}
\label{rgs_letg}
\end{figure*}

Simultaneously with the 2004 June 21 \xmm\ observation, \pgt\ was
also observed with the Low Energy Transmission Grating (LETG) on
board the {\it Chandra} X-ray observatory. The LETG observation of
$\sim 45$~ks has made use of the ACIS CCDs as the detector. We
retrieved the data of this observation from the {\it Chandra
archive} and reduced it using CIAO 3.2.1 and CALDB version 3.01,
according to the updated CIAO threads.  We have combined the +1
and -1 orders of the LETG spectrum using a weighted mean and the
combined spectrum is represented in Figure~\ref{rgs_letg} by a
black line. The simultaneous RGS observation (the red data in
Figure~\ref{2rgs}) is shown in red in Figure~\ref{rgs_letg}.

Although the simultaneous data from the two X-ray observatories
are consistent overall, they differ in many details. For example,
the RGS data between 7 to 9 \AA\ show several emission-like
features which the LETG data do not. Also, around 16.4 \AA\ the
LETG data show absorption-like features which are not present in
the RGS data. These differences are consistent to within about
3$\sigma$ and are probably a result of the poor S/N of the
observations

After the first $\sim$\,45\,ks LETG observation of \pgt\ on 2004
June 21, which is described above, there were two more observations
in consecutive orbits of the {\it Chandra} observatory. The second
$\sim 45$\,ks observation took place on 2004 June 23 and the third
observation was on 2004 June 25. These data are not presented here, but
their spectra is in overall agreement with the first LETG observation,
{\it except} that the flux level in the last two observations was
twice that of the first observation, i.e., during a period of $\sim
2$ days the flux level doubled. This is somewhat unexpected for
a source that had retained its flux level of three years earlier
(see Figure~\ref{2rgs}). Besides the change in flux between the three
LETG observations, there is also a change in the absorption features
seen between the first observation and the other two.  Some of these
features have disappeared between the first low-flux observation and
the high-flux observations taken 2 days later, while other features
seems to appear. The absorption features interpreted by Reeves et
al. (2005) as evidence for sub-c gravitational infall are seen only
in the second observation and not in the first or third ones. The
fact that, again, absorption features are not reproduced in different
spectra is rather confusing. If these lines are statistically
significant (see Reeves et al. 2005) then they represent a transient flow.

\section{Summary and Conclusions}

We claim in Kaspi \& Behar (2006) that an outflowing absorber
at a velocity of 3000 \kms\ fits the first (2001) RGS data of
\pgt\ better than a 24000 \kms\ model. Admittedly though, the poor
S/N of those data can tolerate more than one interpretation.

A second RGS observation taken three years after the first
observation shows general consistency with the first observation,
but differs in important details of the absorption lines relevant
to the outflow. Some features that appear in the first RGS
observation disappear in the second one and vice versa. This could
be a result of either short-time variability of the absorber
(almost impossible to prove or refute) or the poor S/N of the
data. Even more confusing is the fact that {\it simultaneous}
observations of \pgt\ with RGS and LETG produce spectra that are
partially incompatible in their absorption lines. This
significantly reduces our confidence in the existence of the
absorption lines and even more so in their identification. The
poor S/N of the data calls for extra caution and careful modeling.

The three LETG observations indicate that the continuum source
changes on a timescale of days. If the discrete features seen in
these spectra are real, they too vary on short timescales. With
the loss of the high-resolution X-ray spectrometer (XRS) on board
{\it Astro-E2}, a very long observation of a good, bright UHV-wind
source with {\it Chandra} or \xmm\ gratings remains as the most
viable approach toward testing what we feel is still a putative
phenomenon of high velocity outflow.

\section*{Acknowledgments}

This research was supported by the Israel Science Foundation (grant no.
28/03), and by a Zeff fellowship to S.K.

\end{document}